\documentclass[aps,superscriptaddress,showpacs,nofootinbib, 10pt]{revtex4-1}
\usepackage{mathrsfs}
\usepackage{bigints}
\usepackage{latexsym,bm}
\usepackage{graphicx}
\usepackage{indentfirst}
\usepackage{slashed}
\usepackage{amssymb}
\usepackage{amsmath} 
\usepackage{bbm}
\usepackage{color}
\usepackage[dvipsnames]{xcolor}
\usepackage{epsfig}
\usepackage[titletoc]{appendix}
\usepackage{multirow}%
\usepackage{rotating}
\usepackage{epstopdf}
\usepackage{extarrows}%add text under or on the arrows
\usepackage{tabularx}
\usepackage{soul} % use this (many fancier options)
\usepackage{xcolor}
\usepackage{lipsum}
\usepackage[colorlinks=true,allcolors=BlueViolet,hyperfootnotes=false]{hyperref}

\usepackage[utf8]{inputenc}

\newcommand{\be}{\begin{equation}} \newcommand{\ee}{\end{equation}}
\newcommand{\ba}{\begin{array}{c}} \newcommand{\ea}{\end{array}}
\newcommand{\bea}{\begin{eqnarray}} \newcommand{\eea}{\end{eqnarray}}

\begin{document}
\title{\Large Neutrino and antineutrino charged-current multi-nucleon cross sections revisited}

\author{J. E. Sobczyk}
\affiliation{Institut f\"ur Kernphysik and PRISMA+ Cluster of 
Excellence, Johannes Gutenberg-Universit\"at Mainz, 55128 Mainz, Germany}

\author{J. Nieves}
\affiliation{Instituto de F\'{\i}sica Corpuscular (centro mixto CSIC-UV), Institutos de Investigaci\'on de Paterna,
Apartado 22085, 46071, Valencia, Spain}

\date{\today}

\begin{abstract}
 In this work we improve on several aspects of the  computation of the (anti-)neutrino charged-current multi-nucleon cross section carried out in  Phys.Rev.C 83 (2011) 045501 and Phys.Rev.C 102 (2020) 024601. Most importantly, we implement a consistent treatment of the nucleon self-energy in the $W^\pm N\to N'\pi$ amplitude entering the definition of the two-particle two-hole (2p2h) cross-section, and estimate the source of uncertainty of our model due to a simplified treatment of the $\Delta$ self-energy.
 Our new predictions are around 20-40\% higher than previously. We show comparisons for the inclusive lepton double-differential cross sections, with no pions in the final state,  measured by  MiniBooNE on carbon and by T2K on carbon and oxygen.  In all cases, we  find an excellent reproduction of the experiments, and in particular,  the neutrino MiniBooNE data is now well described without requiring a global 90\% re-scaling of the flux.  In addition, we  take the opportunity of this revision to discuss in  detail several important issues of the calculation of the 2p2h cross section, delving into the microscopic dynamics of the multi-nucleon mechanisms.  The improved treatment presented in this work provides realistic first-step  emitted two-nucleon final state momentum configurations, beyond the approximation of phase-space distributions. 

\end{abstract}
%\keywords{...}

\maketitle
\section{Introduction}

The ambitious experimental programs to precisely determine neutrino properties, test the three-generation paradigm, establish the order of mass eigenstates and investigate leptonic CP violation are underway~\cite{DUNE:2020jqi,Hyper-KamiokandeProto-:2015xww}. The success of these endeavours will require an excellent control over the systematic uncertainties, and therefore a solid microscopic understanding of neutrino-nucleus interactions~\cite{Alvarez-Ruso:2017oui}. 
Neutrino and antineutrino beams used in long-baseline oscillation experiments are not monochromatic making the experiments sensitive to all contributing mechanisms up to the energies of the order of 1 GeV. Therefore, the reconstruction of the incident neutrino energy requires a consistent framework in which all these relevant interaction channels are included, not only at the level of inclusive cross-sections, but also giving access to exclusive final states~\cite{Filali:2024vpy}. 
This modelling is a challenging task which encompasses a theoretical description of the so-called primary vertex of interaction, in which hadrons are produced inside of a nucleus, as well as the intra-nuclear cascade which can lead to a different final distribution of particles.

As a prominent example, the CC0$\pi$ cross section, charge-current (CC) driven (anti-)neutrino reaction with no pions in the final state, reported by experiments, requires also accounting for real pion production in the first step, after which the final pion gets absorbed in the nuclear medium. The so-called CC quasielastic-like (CCQE-like) cross section can be obtained after Monte Carlo correcting the CC$0\pi$ one for those events where CC pion production was followed by pion absorption. However, it still does not correspond to the pure quasielastic (QE) process since multinucleon mechanisms, where the gauge boson $W^\pm$ is absorbed by a pair of interacting nucleons,  should be accounted for.  Mismatching the signal coming from these two reaction mechanisms would lead to a bias in the (anti-)neutrino energy reconstruction \cite{Nieves:2012yz}.

In this paper we follow a theoretical framework which consistently describes (anti-)neutrino QE, multi-nucleon knock-out  and pion production processes~\cite{Nieves:2004wx, Nieves:2011pp}. The same theoretical approach has been widely used to model the pion propagation in the nuclear medium, accounting for pion absorption, charge exchange and production mechanisms. 
Actually, the scheme is an extension of the framework developed by the Valencia group, successfully used for many different nuclear reactions at intermediate energies. Importantly, these works  include inclusive scattering of photons~\cite{Carrasco:1989vq}, electrons~\cite{Gil:1997bm} and pions from nuclei~\cite{Salcedo:1987md,Garcia-Recio:1989hyf,Nieves:1993ev,Nieves:1991ye}, where multi-nucleon mechanisms played a crucial role. 
The fundamental object in the application of this approach for (anti-)neutrinos is the imaginary part of the $W$ boson self-energy inside of the nuclear medium. The scheme presented in Ref.~\cite{Nieves:2011pp} amounts to perform a  many-body expansion in which  the relevant gauge boson absorption modes are  systematically incorporated: absorption by one nucleon (QE), or a pair of nucleons  or even three nucleon mechanisms, real and virtual meson ($\pi$, $\rho$)  production, excitation of  $\Delta$ or higher-energy resonance degrees of freedom, etc. In addition, nuclear effects such as RPA collective excitations  or short range correlations (SRC) are also taken into account. 
In particular, the nucleon-nucleon interaction, when modeling the excitation of a two-particle two-hole (2p2h) nuclear component, goes far beyond the simple pion exchange, commonly  used in the literature, and which is shown here to be residual for intermediate energy and momentum transfers. It is used the so-called induced interaction~\cite{Oset:1981ih,Ericson:1988gk, Nieves:1996fx}, which consists of longitudinal and transverse channels opened by pion and rho exchanges, respectively, modulated by SRC and RPA re-summed to account for collective effects. This is one of the pillars of the model  developed by the Valencia group  to study many different nuclear reactions at intermediate energies.

We will focus in this work on the 2p2h mechanism.  It stirs a lot of attention, since already for electron scattering, the so-called ``dip'' region is often underestimated by theoretical modeling. Several approaches have been developed to account for this nuclear process, like the SuSA model~\cite{Megias:2016fjk,Megias:2016lke,Megias:2017cuh,Ivanov:2018nlm}, relativistic mean-field~\cite{Gonzalez-Jimenez:2019qhq,Martinez-Consentino:2021vcs,Martinez-Consentino:2023dbt,Martinez-Consentino:2023hcx}, spectral functions~\cite{Lovato:2023khk} or short-time approximation~\cite{Pastore2020}. Currently, a lot of effort is  devoted to deliver predictions at the semi-exclusive level, i.e. accounting for the final nucleons' distributions. As already mentioned, it is a crucial observable for neutrino oscillation experiments. While some theoretical models have intrinsic problems to predict directly exclusive distributions, in our framework this can be achieved in a straightforward way as it was shown in Ref.~\cite{Sobczyk:2020dkn}.

Here, we improve on several aspects of the  computation of the (anti-)neutrino CC multi-nucleon cross section carried out in Ref.~\cite{Sobczyk:2020dkn}.  Most importantly, we implement a correct treatment of nucleon self-energy in the $W^\pm N\to N'\pi$ vertex which enters in the calculation. This consistent description leads to 2p2h strengths around 40\% higher  than  those obtained in the  previous works of Refs.~\cite{Sobczyk:2020dkn, Nieves:2011pp}.
We have also identified the largest source of theoretical uncertainty in the 2p2h cross sections presented in this work and  have displayed error bands in all our predictions.
In addition, we  take the opportunity of this revision to discuss in  detail several important issues of the calculation of the 2p2h cross section, delving into the microscopic dynamics of the multi-nucleon mechanisms. 

This work is organized as follows. In Sect.~\ref{sec:micro}, we review the main elements of the calculation of the CCQE-like cross section carried out in Refs.~\cite{Nieves:2011pp,Sobczyk:2020dkn}. We first briefly describe the model for the pion production on the nucleon, paying special attention to recent developments not considered in \cite{Nieves:2011pp}. Subsect.~\ref{sec:int} is devoted to the induced spin-isospin $NN, N\Delta, \Delta\Delta$ interactions inside of a nuclear medium.  In Subsec.~\ref{sec:further}, further physical effects are presented, together with a correct nucleon self-energy treatment in the in-medium $W^\pm NN$ vertex and the identification of the largest source of theoretical uncertainty in this calculation of the 2p2h cross sections. Next, in Sect.~\ref{sec:results}, we show predictions for the MiniBooNE carbon CCQE-like and T2K carbon and oxygen CC0$\pi$ lepton double-differential cross sections. 
We also compare with some published distributions obtained within the Lyon (\cite{Martini:2009uj, Martini:2011wp}) and previous Valencia (\cite{Sobczyk:2020dkn, Nieves:2011yp}) schemes. Finally, the main conclusions of this work are collected in Sect.~\ref{sec:concl}.

\section{Microscopic calculation of 2p2h cross section}
\label{sec:micro}

\begin{figure}[t]
    \centering
    \includegraphics[height=.22\textwidth]{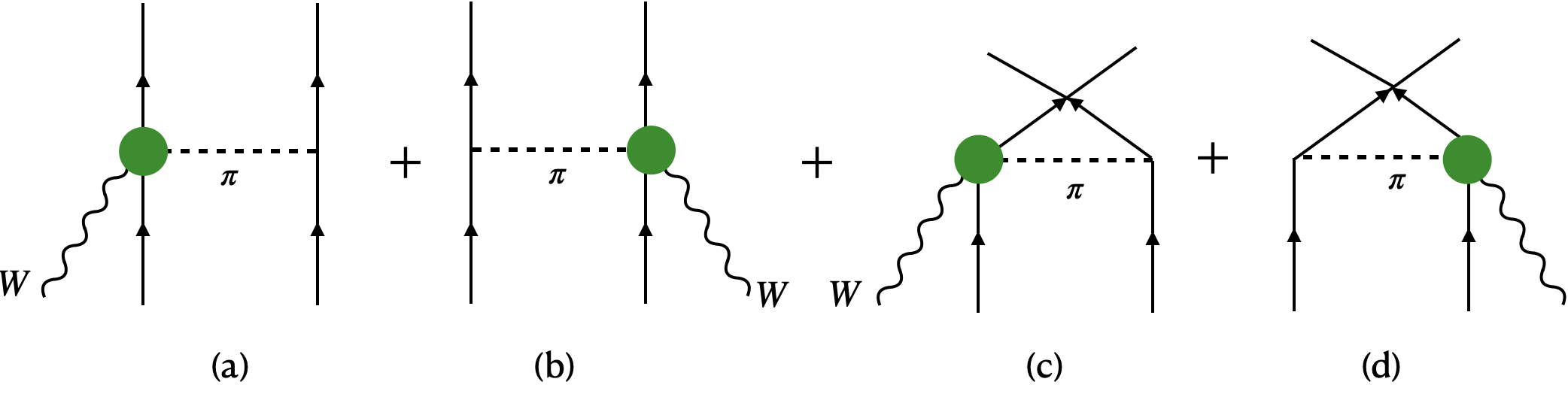}\\\vspace{1.5cm}
    \includegraphics[height=.3\textwidth]{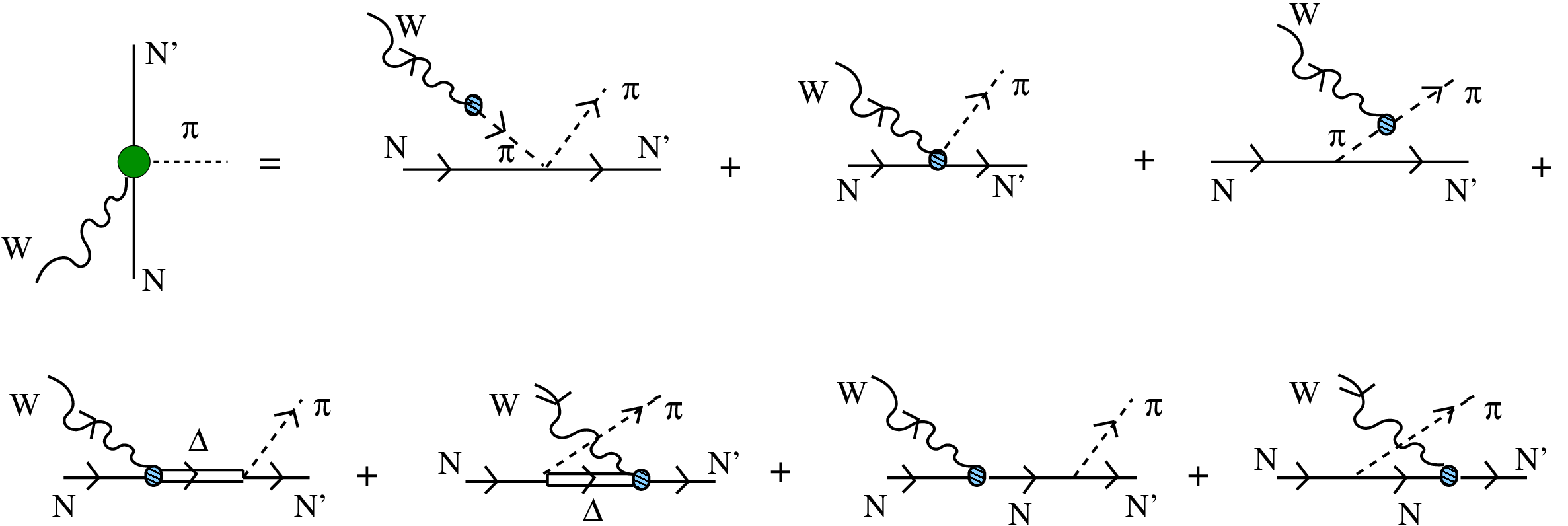}
    \caption{Top: four diagrams contributing to 2p2h  constructed from the $WN\to N'\pi$ amplitude (green circle-box): pion production from the first (a) and second (b) of the initial nucleons and the corresponding anti-symmetrized terms (c) and (d), which are obtained from the previous ones by crossing the final nucleons.  
     Bottom: Model for the $WN\to N'\pi$ amplitude, comprised of seven Feynman diagrams: pion pole (PP), contact term (CT), pion in flight (PF), Delta pole ($\Delta$P) and cross Delta pole (C$\Delta$P), and nucleon pole  (NP) and cross nucleon pole (CNP). The light blue circle-boxes in the diagrams stand for the weak transitions and include phenomenological form-factors.}
    \label{fig:curr}
\end{figure}

The basic dynamical ingredient driving the 2p2h mechanism is the (anti-)neutrino pion production off nucleons (see top panel of Fig.~\ref{fig:curr}). The $W^\pm N\to N'\pi$ amplitude,  represented by the green dot, is comprised of seven contributions depicted at the bottom panel of the figure.  The model developed in \cite{Hernandez:2007qq} contains the excitation of the $\Delta(1232)$ resonance and non-resonant background, required by chiral symmetry. It was further improved in  Ref.~\cite{Alvarez-Ruso:2015eva} by unitarizing the leading vector and axial multi-poles applying Watson's theorem.  
The most recent version of the model was derived in Ref.~\cite{Hernandez:2016yfb}, where the $\Delta$ propagator was changed from the Rarita-Schwinger form to the pure spin-$3/2$ projector operator and consistent couplings~\cite{Pascalutsa:2000kd} were employed.\footnote{In Ref.~\cite{Hernandez:2016yfb}, there is nevertheless a very small local contribution stemming from a possible $1/2$-spin part in the off-shell $\Delta$ propagator that is fitted to the $\nu_\mu n \to \mu^- n \pi^+$ cross section data.} 
As a consequence, a $p^2_\Delta/M^2_\Delta$ local factor, which is equal to one at the $\Delta-$peak, appears in the $\Delta$ propagator, which now reads
\begin{equation}
G_{\mu\nu}(p_\Delta) = \frac{p^2_\Delta}{M^2_\Delta} \frac{P_{\mu\nu}^{\frac32}(p_\Delta)}{p_\Delta^2-M_\Delta^2+i M_\Delta \Gamma_\Delta}\ ,  \quad P_{\mu\nu}^{\frac32}(p_\Delta) = -(\slashed{p}_\Delta+M_\Delta) \bigg[ g_{\mu\nu} - \frac{1}{3} \gamma_\mu \gamma_\nu - \frac{1}{3p_\Delta^2} (\slashed{p}_\Delta \gamma_\mu p_{\Delta\nu}+p_{\Delta\mu} \gamma_\nu \slashed{p}_\Delta)   \bigg] \label{eq:factor}
\end{equation}
These improvements lead to  a substantially  enhanced $\nu_\mu n \to \mu^- n \pi^+$ cross section, which is now quite well described\footnote{The factor $p^2_\Delta/M^2_\Delta$ in Eq.~\eqref{eq:factor} reduces considerably the contribution of the $\Delta-$cross term of the amplitude in all channels, especially in the $\nu_\mu n \to \mu^- n \pi^+$ one. }, 
together with the rest of available data from the Argonne and Brookhaven bubble-chamber experiments. Moreover, the resulting model leads to an accurate reproduction of the Watson's unitary theorem in the dominant $P_{33}-$multipoles without including any further sizable complex phase~\cite{Hernandez:2016yfb}, which is quite reassuring from the theoretical perspective. Finally, the complete  model of Ref.~\cite{Hernandez:2016yfb} also includes the contribution from the $N^*(1520)$ resonance which is the only resonance, apart from the $\Delta(1232)$, that gives a significant contribution in the few GeV neutrino energy region \cite{Leitner:2008wx}.
It is interesting to note that the model was  successfully  confronted to experimental total and differential cross section for pion production off nucleons induced by both real and  and virtual photons~\cite{Hernandez:2016yfb,Sobczyk:2018ghy}. In the latter case,  the theoretical predictions for polarized-electron reactions were also compared to data, as a further test of the vector content of the model. 
Moreover, an exhaustive comparison of the pion angular distributions obtained within the scheme of Ref.~\cite{Hernandez:2016yfb} and  the  Argonne-Osaka DCC (dynamical coupled-channel) model~\cite{Matsuyama:2006rp,Kamano:2013iva, Nakamura:2015rta} was carried out in \cite{Sobczyk:2018ghy}. Given the high degree of complexity of the DCC approach, it is really remarkable that the bulk of its predictions for electroweak production of pions in the $\Delta$ region could be reproduced, with a reasonable accuracy, by the simpler  model of Ref.~\cite{Hernandez:2016yfb}.

For simplicity, in the 2p2h calculations below, we will neglect the weak excitation of the $D_{13 }$ $N^*(1520)$ resonance  which is negligible for MiniBooNE and T2K experiments, and so we employ only the diagrams depicted in the lower panel of Fig.~\ref{fig:curr}. Furthermore, we will strictly consider the $\Delta(1232)$ propagator of Eq.~\eqref{eq:factor}, neglecting any spin $1/2$ off-shell term  and the ($W_{\pi N},q^2$)-dependent unitarity relative phases between background and resonant terms deduced from Watson's theorem since  they become very small once consistent $\Delta$ couplings are used (Fig.~5 of Ref.~\cite{Hernandez:2016yfb}).

\subsection{ Nuclear 2p2h cross section: meson exchange, short range correlations  and RPA} 
\label{sec:int}
\begin{figure}[h]
\centering
\includegraphics[height=.25\textwidth]{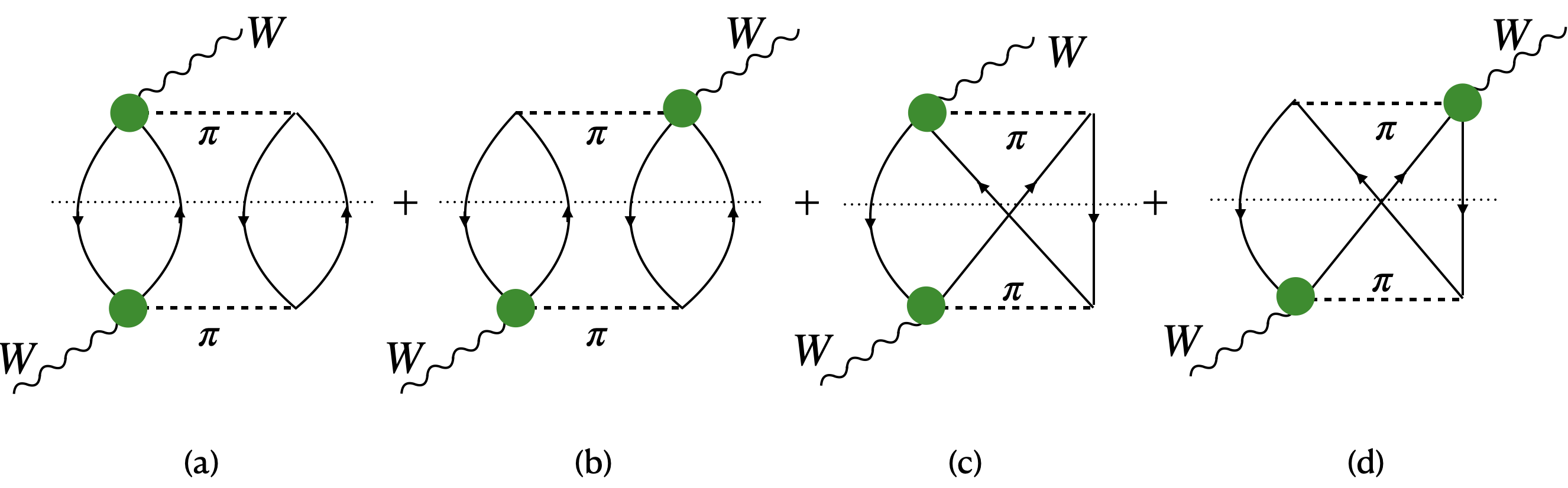}
\caption{2p2h many-body contributions to the $W$ self-energy driven only by pion exchange. The green circle-boxes account for the seven $W^\pm N\to N'\pi $ Feynman diagrams shown in the two bottom rows of Fig.~\ref{fig:curr}. The horizontal dotted lines show the cuts that put on the mass-shell the 2p2h excitations and that account for $WNN\to NN$ absorption inside of the nuclear medium. }
\label{fig2:w_se}
\end{figure}

In many-body quantum field theory~\cite{Fetter1971,Mattuck:1976xt}, the diagrams in the upper panel of Fig.~\ref{fig:curr} represent the $WNN\to NN$ absorption reaction inside the  nuclear medium. They can be equivalently expressed as the imaginary parts of the $W$ self-energy diagrams depicted in Fig.~\ref{fig2:w_se}, where the initial and outgoing nucleons are put on-shell by the Cutkosky cuts shown in the figure. 
The sum of the modulus squared of each of the four 2p2h mechanisms  depicted in Fig.~\ref{fig:curr} leads to the many-body diagram (a) of Fig.~\ref{fig2:w_se}, while the interference between (direct and crossed) mechanisms where the pion is produced in the
first or in the second of the initial nucleons is accounted by the many-body diagram (b). These are the contributions considered in our approach. The (c) and (d) mechanisms of Fig.~\ref{fig2:w_se} account for the direct-exchange interference~\cite{RuizSimo:2016rtu}, i.e., contributions stemming from the interference between the direct   and the antisymmetrized diagrams of Fig.~\ref{fig:curr}. We will further comment on the  direct-exchange interference contributions below. 

Our approach is based on the local density approximation\footnote{We would like to point out that the local Fermi gas  provides  nucleon momentum distributions much more realistic than those obtained in the global Fermi gas model, as illustrated in Figs.~3 and 4 of Refs.~\cite{Bourguille:2020bvw} and \cite{Alvarez-Ruso:2014bla}, respectively. } (LDA), following the density power counting (number of hole lines). In Sec. IIA of~\cite{Sobczyk:2020dkn}, we explained in detail, within the adopted scheme, the  many-body diagrams that account for  initial and final state correlations and meson exchange current contributions. There we also discussed the power density counting of each of the latter many-body mechanisms.   

The key feature of the approach lies in employing a widely tested effective interaction in the nuclear medium between $NN, N\Delta, \Delta\Delta$, i.e. in the diagrams driven by the $\Delta$P, C$\Delta$P, NP and CNP vertices (green dots) in Fig.~\ref{fig2:w_se}. Such isovector baryon-baryon interaction  accounts for both the spin longitudinal and transverse channels, going beyond the one pion exchange (OPE)  potential, and includes short and long range correlations.
We note that this differentiates our approach from many calculations of the 2p2h cross section~\cite{Megias:2017cuh, RuizSimo:2016rtu,Megias:2016fjk,Lovato:2023khk,Martinez-Consentino:2023dbt} which are based only on OPE.

The longitudinal and transverse channels opened by $\pi$ and $\rho$ exchanges, respectively,  are firstly supplemented by the SRC $g_l'(k)$, $g_t'(k)$ terms, which account for the dynamics at shorter distances and effectively prevent nucleons from getting close to each other. The SRC Landau-Migdal parameters $g'_{l/t}(k)$ inside of the nuclear medium have a smooth dependence on the carrier's four-momentum $k$~\cite{Oset:1981ih,Oset:1987re,GarciaRecio:1987hwz}, which has been recently re-derived in Ref.~\cite{Sobczyk:2020dkn}. Thus in momentum space, the effective interaction reads for the $NN\to NN$ case
\begin{equation}
\begin{split}
   &V_l(k) = \frac{f_{\pi NN}^2}{m_\pi^2}\left[F_{\pi}^2(k^2)\, \frac{\vec{k}^2}{k^2-m_\pi^2+i\epsilon} + g'_l(k)\right](\vec{\tau}_1 \cdot \vec{\tau}_2) \, \vec{\sigma}_1^i \vec{\sigma}_2^j\,  \hat{k}^i\hat{k}^j\; ,  \\
   &V_t(k) = \frac{f_{\pi NN}^2}{m_\pi^2}\left[C_\rho F_{\rho}^2(k^2)\, \frac{\vec{k}^2}{k^2-m_\rho^2+i\epsilon} + g'_t(k) \right](\vec{\tau}_1 \cdot \vec{\tau}_2) \, \vec{\sigma}_1^i \vec{\sigma}_2^j\,  (\delta^{ij}-\hat{k}^i\hat{k}^j),
   \label{eq:vlt}
\end{split}
\end{equation}
with $\hat{k} =\vec{k}/|\vec{k}\,|$ the unitary three momentum of the virtual pion or rho mesons, $\sigma_i$ and $\tau_i$ ($i=1,2$), Pauli matrices acting on the spin and isospin nucleon degrees of freedom of each nucleon, respectively, $C_\rho=2$, $f^2_{\pi NN}/4\pi=0.08$ and  the form factors ($m_\pi =139\ \text{MeV}, m_\rho =770\ \text{MeV}$)
\begin{equation}
 F_{\pi/\rho}(k^2)=\frac{\Lambda_{\pi/\rho}^2-m_{\pi/\rho}^2}{\Lambda_{\pi/\rho}^2-k^2}\, , \quad k^2=(k^0)^2-\vec{k}^{\,2}\, , \quad \Lambda_\pi=1200\ \text{MeV}, \quad \Lambda_\rho=2500\ \text{MeV} 
\end{equation}
The $N\Delta$ and $\Delta\Delta$ interactions are obtained by replacing $\vec{\sigma} \to \vec{S} $,
$\vec{\tau} \to \vec{T} $, where $\vec{S},\vec{T} $ are the spin,
isospin $N\Delta$ transition operators~\cite{Oset:1981ih,Ericson:1988gk} and $f_{\pi NN}\to
f^*_{\pi N\Delta}=2.13~f_{\pi NN}$, for any $\Delta$ which replaces a nucleon. 

For moderate momenta, $V_l(k)$ is suppressed due to cancellations between OPE and the SRC $g'_l(k)$ term\footnote{The SRC term $g'_l(k)$ is not the same one that is fitted in the free space, since SRC are modified inside of the nuclear medium. Values  around 0.6-0.7, fitted for the Valencia group in the 1980's and 1990's, led to successful data descriptions of inclusive photo-, electron- and pion-nuclear reactions, of pionic atom levels, and of many other processes sensitive to the (pseudoscalar \& vector)-isovector part of the $NN, N\Delta, \Delta\Delta$ interactions, which involved moderate momenta around 300-400 MeV. This range of values for $g'_l$ is also consistent with the results of Ref.~\cite{Speth:1980kw} from calculations of
nuclear electric and magnetic moments, transition probabilities,
and giant electric and magnetic multipole resonances.  In this regime, we can see that the SRC greatly cancels out the OPE interaction,  since 
\begin{equation}
  \vec{k}^2 D_\pi(k^2)\sim -1 + [m_\pi^2-(k^0)^2]/\vec{k}^2+\cdots  \label{eq:vlcont}
\end{equation}
 and as result, the longitudinal channel of the interaction turns out to be largely suppressed as compared to the OPE  (see left plot of Fig.~5 of Ref.~\cite{Sobczyk:2020dkn}).} and so the transverse channel of the isovector part of the $NN$, $N\Delta$, $\Delta\Delta$ interaction plays a crucial role in many nuclear reactions at intermediate energies. 
We note that the (pseudoscalar and vector)-isovector part of the $NN$ interaction is of little importance to describe the bulk of the structure of closed-shell isospin symmetric nuclei, which is mostly given by scalar-isoscalar and vector-isoscalar exchanges (e.g. Quantum hadrodynamics~\cite{Serot:1997xg}  begins with effective $\sigma$ and $\omega$ meson exchanges). However, this piece of the interaction becomes extremely important when external probes (pions, real and virtual photons or $W^\pm$ and $Z^0$ bosons) excite these spin-isospin components of the nuclei, and it becomes really essential when the transferred energies make it possible to excite the  $\Delta$ resonance. Thus, in these situations, the nuclear responses are in great extent determined by these degrees of freedom that otherwise have little relevance in the nuclear structure.

\label{sec:induced}
\begin{figure}[t]
\centering
\includegraphics[height=.3\textwidth]{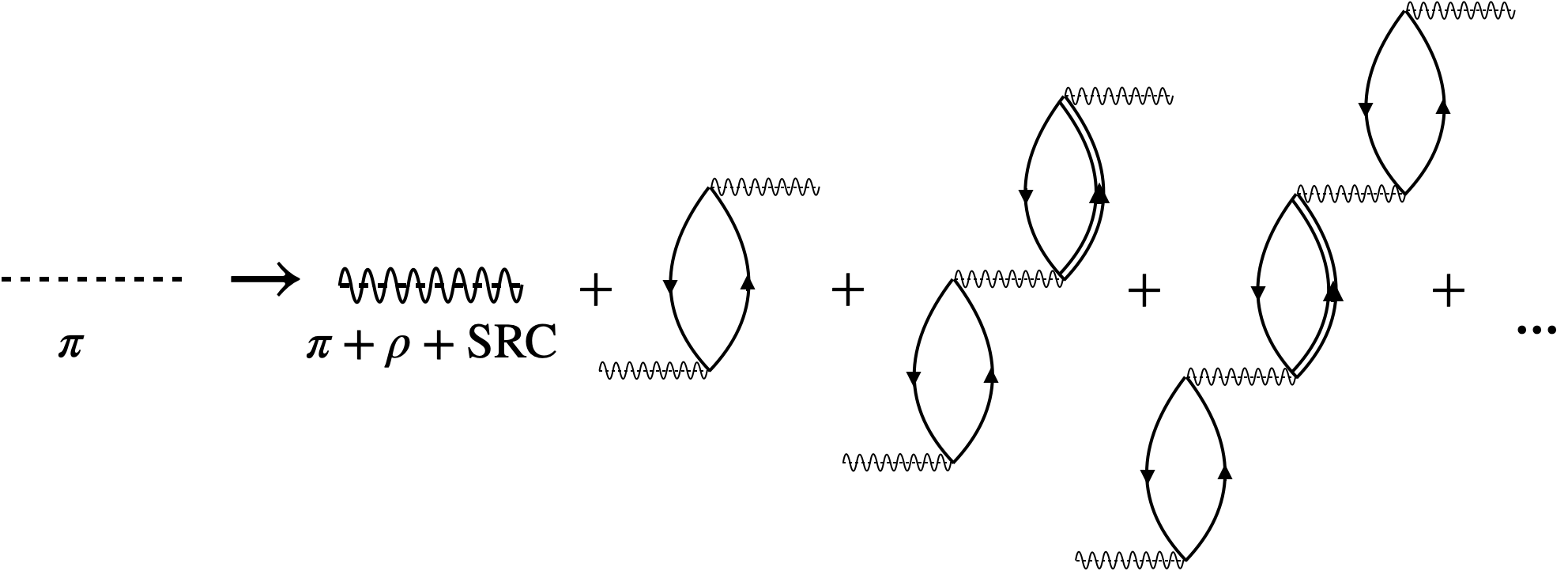}
\caption{Diagrammatic representation of the RPA series of 1p1h and $\Delta$h excitations, driven by the effective $\pi+\rho+ {\rm SRC}$ interaction.}
\label{fig:interaction}
\end{figure}

The $\pi+\rho+{\rm SRC}$ model for the $NN$, $N\Delta$, $\Delta\Delta$ medium interactions is further improved by including nuclear collective polarization corrections, as diagrammatically shown in  Fig.~\ref{fig:interaction}.  Hence, ultimately the picture for the spin-isospin dependent part of the interaction accounts also for the propagation of the mesons through the medium in the RPA sense, producing indistinctly 1p1h and $\Delta$-hole ($\Delta$h)  nuclear excitations in the whole nucleus.\footnote{This has an obvious resemblance with the properties of the instantaneous Coulomb interaction in an electron gas. One electron inside of an electron gas polarizes the medium in such a way that, in a region around it, the negative charges are slightly displaced away from the electron, leaving behind the background charge of the positive ions. As a consequence of this polarization, the photon acquires an effective mass, which leads to the screening of the Coulomb interaction~\cite{Fetter1971}.} 
Thus, one finds two (longitudinal and transverse) geometrical series, which do not interfere, and that can be easily summed up leading to the induced spin-isospin interaction, which
derivation can be found e.g. in Refs. ~\cite{Oset:1981ih,Ericson:1988gk} or more recently in the Section 2.2 of Ref.~\cite{Nieves:1996fx}. Thus in the many body diagrams of Fig.~\ref{fig2:w_se}, one should replace $ V_\pi(k)$ (OPE) by the induced spin-isospin interaction in all interaction lines that emerge from the direct and cross $\Delta(1232)$ and nucleon pole terms ($\Delta$P, C$\Delta$P, NP and CNP). Pions in the PP, PF and CT vertices should not be modified because those mechanisms  require an actual pion.

This  induced spin-isospin interaction has been satisfactorily used for many different nuclear reactions at intermediate energies by Prof. Oset's  Valencia group~\cite{Oset:1979bi,Oset:1981ih,Oset:1985uti,Salcedo:1987md,Oset:1987re,GarciaRecio:1987hwz,Carrasco:1991we,FernandezdeCordoba:1992ky,FernandezdeCordoba:1993az,Garcia-Recio:1991ocp, FernandezdeCordoba:1991wf, Gil:1997bm,Carrasco:1989vq,Nieves:1993ev, Nieves:1991ye}, and that importantly  include inclusive scattering of photons~\cite{Carrasco:1989vq}, electrons~\cite{Gil:1997bm} and pions from nuclei~\cite{Nieves:1993ev,Nieves:1991ye}, where multinucleon mechanisms play a crucial role. 
In all these works, the many-body exchange diagrams of panels (c) and (d) of Fig.~\ref{fig2:w_se} had never been considered. They were discussed in the context of the calculation of the $\Delta$ self-energy in the pioneering work of Ref.~\cite{Oset:1987re} (see text related to Fig. 4 of that reference). Firstly, such contributions were found to produce corrections of the order of $25\%$, which agrees with the claims made in Ref.~\cite{RuizSimo:2016rtu} using only OPE. Doubts emerged, however, due to to some undesirable cancellation between $g'$-contributions in the direct and exchange diagrams.  It was argued in \cite{Oset:1987re} that $g’$ is associated to the repulsive forces that arise as a consequence of the antisymmetry of the quarks when two bags overlap, and hence it is the antisymmetry of the quarks what matters and one should not play with the antisymmetry of the nucleons.  In any case, within the scheme adopted by the Valencia group,  such exchange diagrams were always neglected, and the effective parameters, like $g'_l$, $g'_t$ or form-factors, of the in-medium interaction were successfully fitted to pion-nuclear inclusive processes (pionic atom data, elastic and inelastic pion-nucleus scattering and pion absorption in nuclei), since the pion precisely selects the  spin-isospin piece of the  $NN$, $N\Delta$, $\Delta\Delta$ interactions. The subsequent results for real photon- and electro-nuclear processes or CC neutrino reactions are predictions of the model. Including at any later point the exchange contribution, would necessarily  require a re-fit of the effective interactions,  since these contributions are implicitly (effectively) included.\footnote{See also the discussion of Fig.~31 of Ref.~\cite{Carrasco:1989vq}, where the interaction of real photons with nuclei from 100 MeV to 500 MeV is studied.}

\subsection{\it Further physical effects and considerations}
\label{sec:further}

\begin{figure}[t]
\centering
\includegraphics[height=.2\textwidth]{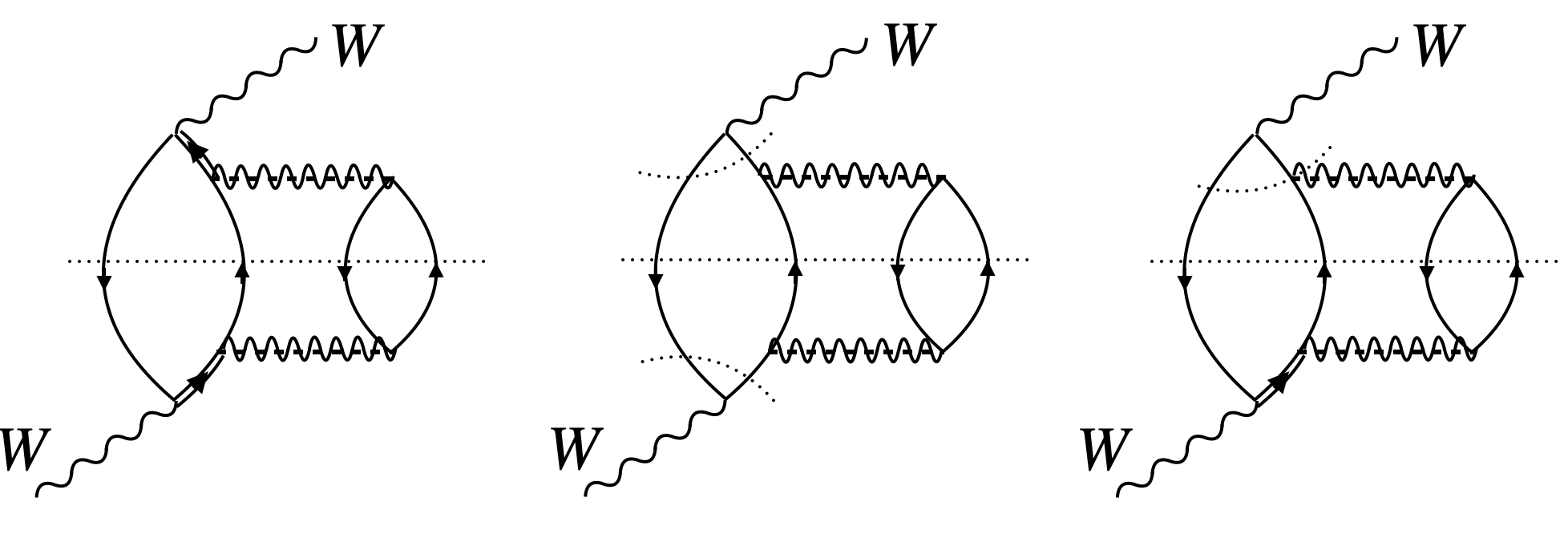}
\caption{ $W$ self-energy many-body diagrams obtained from the diagonal $\Delta$P-$\Delta$P (left), diagonal NP-NP (middle) and off-diagonal $\Delta$P-NP (right) terms of the $WN\to N'\pi$ amplitude. In all diagrams the OPE interaction has been replaced by the full induced interaction of Fig.~\ref{fig:interaction}. Note that the NP-NP diagram can be also seen as a QE contribution where the nucleon propagator of the particle state in the 1p1h excitation is dressed. In all cases the 2p2h cut is depicted, and in the off-diagonal  $\Delta$P-NP and diagonal NP-NP terms, 1p1h cuts are also shown to illustrate that even though the 2p2h excitation is on-shell, one can still put on the mass shell the nucleons cut by the circular dotted line. }
\label{fig:single-pole}
\end{figure}

To compute the 2p2h cross sections we adopt in this paper the scheme of Ref.~\cite{Sobczyk:2020dkn}. 
The latter work improves on some aspects of the  previous work of Ref.~\cite{Nieves:2011pp}, the main motivation being to retrieve the distribution of outgoing nucleons in the primary reaction, which otherwise were distributed according to phase-space in Monte Carlo event generators. 
To this end, we separated the two (2p2h) and three (3p3h)  nucleon contributions driven by the $\Delta$ excitation, which in Ref.~\cite{Nieves:2011pp} were treated together. Moreover, in the description of the in-medium interactions between nucleons and $\Delta$'s, the energy- and momentum-transfer dependence of the SRC $g^\prime_l$ and $g^\prime_t$ is maintained. In addition, the treatment of the weak excitation of the $\Delta$ resonance was also refined, according to the latest model for the $WN\to N'\pi$ amplitude derived in Ref.~\cite{Hernandez:2016yfb}, and it includes  changes in the $\Delta$ propagator (see discussion of Eq.~\eqref{eq:factor})  and in the value at $q^2=0$ of the dominant axial $(C_5^A(q^2))$ $WN\Delta$ form factor, which is in excellent agreement with the value derived from the $\pi N\Delta$ coupling (off-diagonal Goldberger-Treiman relation) and 
around a 20\% larger than that used in Ref.~\cite{Nieves:2011pp}. Finally, we did not average  in \cite{Sobczyk:2020dkn} the initial nucleons' momenta.\footnote{The computation of the 2p2h contribution of the many-body diagrams involving NP terms is subtle, since  when placing the 2p2h excitation on-shell, the nucleon propagator that is part of the $W N\to N'\pi$ amplitude can be still placed on-shell (circular cuts in Fig.~\ref{fig:single-pole}) for a virtual $W$, and thus there exists a single pole in the integration over the hole three momentum.  The correct treatment requires i) to dress the nucleon propagator with a complex selfenergy $\Sigma_N(p)$, which moves the pole to the complex plane out of the integration domain, and ii) to explicitly perform all the nucleon phase-space  integrals avoiding the use of any type of average over the kinematics of the nucleons~\cite{Gil:1997bm}. }

In what follows we enumerate some considerations and additional improvements on the treatment of Ref.~\cite{Sobczyk:2020dkn} implemented in this work.
\begin{itemize}

\item[a)] {\it $\Delta$ self-energy} --- As pointed out in Ref.~\cite{Nieves:2011pp}, the RPA series of $\Delta$h excitations driven by the $\pi+\rho+{\rm SRC}$ spin-isospin interaction\footnote{Actually, one should use the full induced spin-isospin interaction sketched in  Fig.~\ref{fig:interaction}, which amounts to account also for intermediate 1p1h excitations in the RPA series (see for instance Fig.~16 of Ref.~\cite{Gil:1997bm}). The discussion below is not affected by including these corrections.} 
gives rise to an irreducible $W$ self-energy piece and therefore to a contribution to the hadron tensor. The latter depends on the particular Lorentz components of the tensor  which is being evaluated. 
Each of these tensor components are the result of matrix elements of different operators ${\cal O}_i$, selecting either the longitudinal or transverse pieces of the $\pi+\rho+{\rm SRC}$ interaction 
\begin{equation}
    \sum_{\rm spins} \sum_{\rm isospins}\ \sum_{i,j\in WN\Delta}\langle N | {\cal O}_i| \Delta \rangle \langle \Delta | {\cal O}_j| N \rangle 
\end{equation}
which leads to the fact that each of the hadron tensor elements $W^{\mu\nu}_\Delta$ could be modified by a different value of the real part of the $\Delta$ self-energy ${\rm Re}\Sigma^{\rm eff}_\Delta(p_\Delta, q;\rho)$. 

In the study of inclusive cross sections of real photons by nuclei, due to the transverse nature of the  $\gamma N\Delta$ vertex, only $V_t$ is relevant which leads to~\cite{Carrasco:1989vq}
\begin{equation}
  {\rm Re}\Sigma^{\rm eff, t}_\Delta(p_\Delta, q;\rho) =   {\rm Re}\Sigma_\Delta(p_\Delta;\rho) + \frac49\left(\frac{f^*_{\pi N\Delta}}{m_\pi}\right)^2 V_t(q)\rho \label{eq:rpaDelta-t}
\end{equation}
where ${\rm Re}\Sigma_\Delta \sim -50 \rho/\rho_0$ MeV \cite{Oset:1987re,Ericson:1988gk} is referred to the potential felt by the nucleons in the LFG model (bound by $\sim -k_F^2(r)/2M$)\footnote{It means that for $\rho=\rho_0$,  the real part of the $\Delta$-nucleus interaction is around 50 MeV more attractive  that the nucleon mean-field potential.}. 
For momentum transfers smaller than $m_\rho$, $V_t$  is dominated by the SRC Landau-Migdal parameter $g'_t\sim 0.6$ and hence  ${\rm Re}\Sigma^{\rm eff, t}_\Delta(\rho)  \sim +30 \rho/\rho_0$ MeV, at most. Thus the RPA re-summation  provides a natural explanation for why, although the $\Delta-$peak is much wider because of the broadening of the $\Delta$ in the medium, its position hardly changes  in inclusive $\gamma A$ nuclear reactions.\footnote{This can be confirmed in the right panel of Fig.~8.17 of the book T. Ericson and W. Weise "Pions in Nuclei" \cite{Ericson:1988gk}. There, the experimental  $\gamma ^{12}$C total cross section as a function of the laboratory photon energy is compared to $12\times \sigma(\gamma p)$. In the left panel of the same figure is shown the  $\pi ^{12}$C total cross section, for which the $\Delta$ peak is around 50-75 MeV shifted towards lower masses, indicating that in that case the $\Delta$ feels an important attraction in the nuclear medium, as we will see next.}

On the other hand, the interaction of nuclei with external pions selects only the spin longitudinal channel of the $\Delta$h-$\Delta$h interaction in the RPA re-summation, leading to \cite{Oset:1981ih,Nieves:1993ev,Nieves:1991ye} the replacement
\begin{equation}
   {\rm Re}\Sigma^{\rm eff, l}_\Delta(p_\Delta, q;\rho) =   {\rm Re}\Sigma_\Delta(p_\Delta;\rho) + \frac49\left(\frac{f^*_{\pi N\Delta}}{m_\pi}\right)^2 V_l(q)\rho \label{eq:rpaDelta-l}
\end{equation}
where $V_l(q)<0$ and it is small because of  the large cancellation between the OPE piece and the SRC Landau-Migdal parameter $g'_l$. Hence, the effective longitudinal real part of the $\Delta$ self-energy remains more attractive than the result of ${\rm Re}\Sigma_\Delta(p_\Delta;\rho)=-50 \rho/\rho_0$ MeV found in Ref.~\cite{Oset:1987re}. 

When considering the CC neutrino scattering, the weak $W^{\pm}$ gauge boson will probe both longitudinal and transverse operators of the $WN\Delta$ vertex and therefore each tensorial component of the hadron tensor will be dominated by either the longitudinal or the transverse RPA re-summation. A correct treatment requires a detailed non-trivial microscopic evaluation, as it was done in \cite{Nieves:2004wx} for the QE contributions. For that reason in the previous calculations of Refs.~\cite{Nieves:2011pp} and \cite{Sobczyk:2020dkn}, the real part of the position of the $\Delta$-peak was not renormalized in the medium taking
\begin{equation}
  {\rm Re}\Sigma^{\rm eff}_\Delta(p_\Delta, q;\rho) = 0  \label{eq:resig0}
\end{equation}
Here, we will allow ${\rm Re}\Sigma^{\rm eff}_\Delta(p_\Delta, q;\rho)$ to vary in the range $[-50\rho/\rho_0, +30 \rho/\rho_0]$ MeV and the dispersion of CCQE-like predictions will be taken as a systematic error of the calculation. 

\item[b)] {\it Removal energy} --- In the calculation of the 2p2h cross section, we introduce two-nucleon removal energy corrections~\cite{Bourguille:2020bvw}, and the energy transferred to the hadronic system is reduced by the two-nucleon separation energy, 
\begin{equation}
q^0=E_{\nu} - E_{\mu} \to q^0_{\rm eff} =E_{\nu} - E_{\mu} -S_{N_1N_2}, \qquad    S_{N_1N_2} = M'_{A-2}-M_A+m_{N_1}+m_{N_2} 
\end{equation}
with $M_{A}$ and $ M^\prime_{A-2}$ the ground-state  masses of the initial and final nuclei, and $m_{N_i}$ the masses of the ejected nucleons. In the calculations below, we take  approximate average values for  $S_{N_1N_2}$  of around 27 and 22 MeV for carbon and oxygen respectively.

\item[c)] {\it Nucleon self-energy} --- In the previous computations of Refs.~\cite{Nieves:2011pp,Sobczyk:2020dkn} the incoming and outgoing nucleons of  $WN\to N'\pi$ amplitude (entering the 2p2h diagrams) are considered to be moving in a local mean-field potential
\begin{equation}
E(\vec{p};\rho)= \sqrt{M^2+\vec{p}^{\,2}\,}-\frac{k_F^2(r)}{2M}, \qquad k_F(r)=\left[\frac{3\rho(r)}{2\pi^2}\right]^\frac13 
\end{equation}
The constant term, $-k_F^2(r)/2M$, which cancels out in the energy conservation, was not included in the energy of the virtual nucleon that appears in the NP and CNP terms of the $WN\to N'\pi$ amplitude. This is inconsistent because this virtual nucleon should feel the same attraction, neglecting momentum dependent nuclear potential terms. 
In the case of virtual $\Delta$ in $\Delta$P diagram, the term $-k_F^2(r)/2M$ amounted to include effectively a repulsive interaction, since no other real part of the $\Delta$-nucleus potential was included.

This inconsistent description induced smaller 2p2h cross sections. Here, we decide to remove everywhere the constant term $-k_F^2(r)/2M$ from the definition of nucleon energies. It does not affect the overall energy-balance of the reaction (since both the initial and final nucleons are shifted by the same constant term). This choice is also  consistent with the calculation of ${\rm Re}\Sigma_\Delta$ in Ref.~\cite{Oset:1987re}, since the numerical values obtained in that work are referred to the potential felt by the nucleons in the LFG model.

The 2p2h  calculation of Ref.~\cite{Nieves:2011pp}, implemented in some Monte Carlo event generators has the same theoretical inconsistency, which is inherited by the derived works of Refs.~\cite{Nieves:2011yp} and \cite{Nieves:2013fr} on neutrino and antineutrino MiniBooNE double differential  cross sections. All the theoretical developments made in the work on the reconstruction of the neutrino energy and the shape of the CCQE-type total cross section of Ref.~\cite{Nieves:2012yz} are correct, although the quantitative predictions are affected. Finally, the 2p2h cross sections shown in Refs.~\cite{Gran:2013kda,Bourguille:2020bvw} for MINERvA energies and in Ref.~\cite{Bourguille:2020bvw} for the T2K CC0$\pi$, were also evaluated  using the results of  Ref.~\cite{Nieves:2011pp}.
The predictions which we will show in the next section for MiniBooNE and T2K experiments supersede all previous calculations. 

\end{itemize}

\subsection{\it Quasi-elastic contribution}
\label{sec:qe}

The description of CCQE-like cross section requires a consistent calculation of the multinucleon and the QE contributions.  The latter process corresponds to a 1p1h nuclear excitation and here we use the results obtained from the LFG model of Ref.~\cite{Nieves:2004wx}. This QE model  accounts for a correct energy balance using the experimental $Q$ values, the Coulomb distortion of the outgoing charged lepton and medium polarization (RPA) effects. In the  spin-isospin channel, the RPA includes also $\Delta$h degrees of freedom and  it is driven by  the $\pi+\rho+{\rm SRC}$ interaction, leading to the induced interaction shown in Fig.~\ref{fig:interaction}. 
Notably, the RPA re-summation was significantly improved in \cite{Nieves:2004wx} by a correct tensorial treatment of the RPA response function and the inclusion of the scalar-isovector channel of the nucleon-nucleon interaction. In the scheme of Ref.~\cite{Nieves:2004wx}, there exists also a possibility of dressing the nucleon propagators in the nuclear medium, which amounts to work with nucleon spectral functions (SF)~\cite{Benhar:2005dj,Benhar:2013dq,Vagnoni:2017hll}.  

Here, however, the QE cross section will be computed within the RPA model of Ref.~\cite{Nieves:2004wx} without including SF effects. The reason for it is the following.
The hole and particle SFs used in \cite{Nieves:2004wx,Nieves:2017lij,Sobczyk:2017vdy,Sobczyk:2019uej} are obtained from the semi-phenomenological nucleon self-energies developed in Ref.~\cite{FernandezdeCordoba:1991wf}. The real part of self-energy (${\rm Re}\Sigma_N$) acts as an energy-dependent effective potential which changes the in-medium dispersion relation of the nucleons~\cite{Nieves:2017lij}. It largely cancels out between the hole and particle states and can be safely neglected.\footnote{Indeed, it provides a high energy tail in the QE distributions, which was shown in Ref.~\cite{Benhar:2010nx} to have little effect in the reproduction of the MiniBooNE two-dimensional CCQE-like cross section~\cite{AguilarArevalo:2010zc}. } The imaginary part, on the other hand, plays a much more important role. It accounts for $NN$ collisions and it is much larger for particle than for hole states~\cite{FernandezdeCordoba:1991wf}, because of the enhanced available phase-space. The events induced by the imaginary part of the nucleon self-energy account for various collision channels (hadronic states) at the primary vertex. If we are interested in the semi-exclusive cross sections, these events (driven by collisional broadening) in the intranuclear cascade would need to be treated separately from the one body (1p1h) ones in the primary reaction, which are obtained after neglecting  ${\rm Im}\Sigma_N$.  All in all, in the calculation of the 1p1h contribution, we neglect both ${\rm Re}\Sigma_N$ and ${\rm Im}\Sigma_N$, while we maintain RPA corrections and relativistic kinematics.
Within this discussion, we point out that the diagonal  NP-NP  diagram (middle one in Fig.~\ref{fig:single-pole}) can be cast as a 1p1h term, where the particle nucleon is dressed with a complex self-energy $\Sigma_N$ obtained from a single 1p1h insertion. 
Hence, it belongs to the QE calculation using SF and so in our calculations this term is excluded from the sum implicit in the generic 2p2h contribution depicted in the first diagram of Fig.~\ref{fig2:w_se}.

%
%
%
%
%  RESULTS
%
%
%
%

\section{Results}
\label{sec:results}

\begin{figure}[t]
\centering
\makebox[0pt]{\includegraphics[height=.75\textwidth]{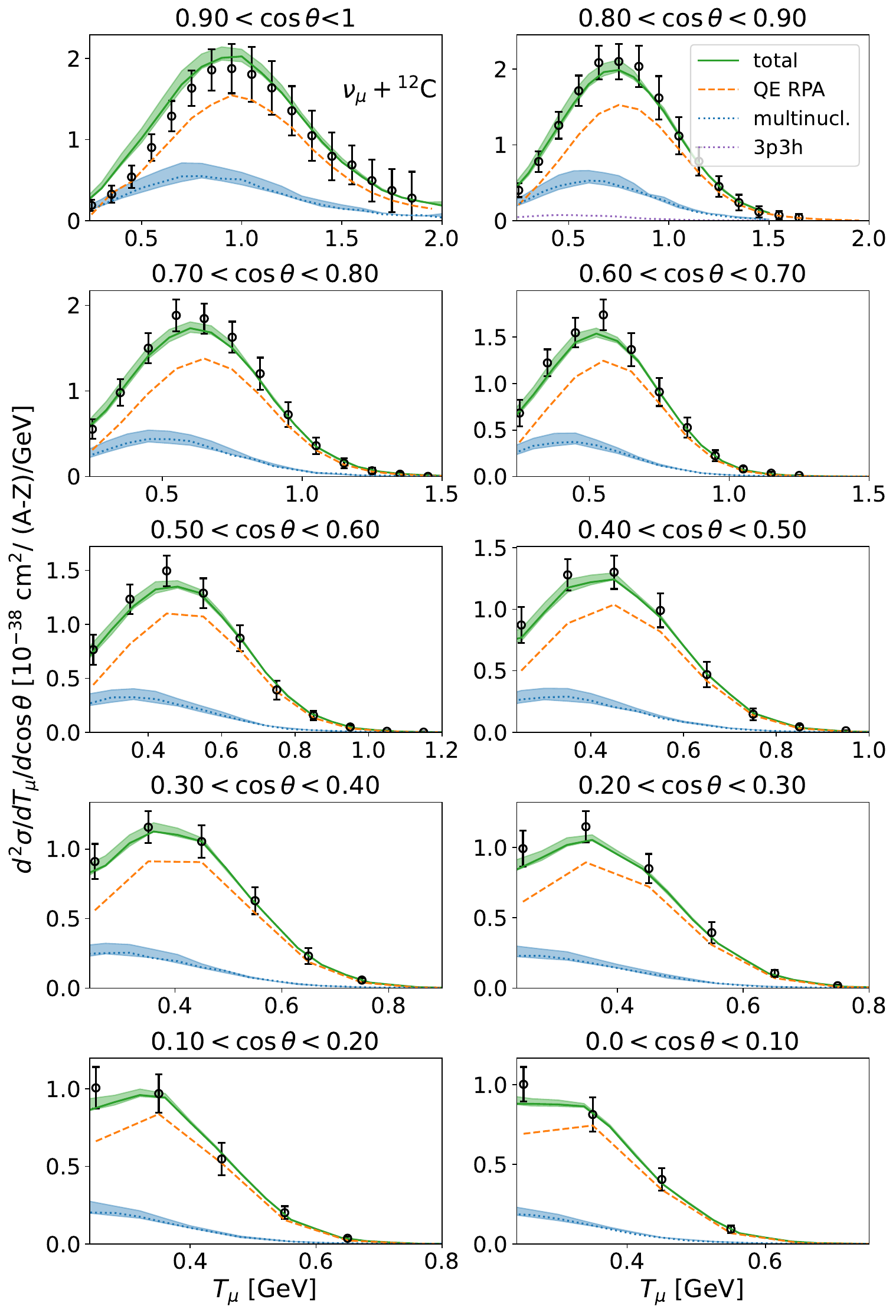}\hspace{0.5cm}\includegraphics[height=.75\textwidth]{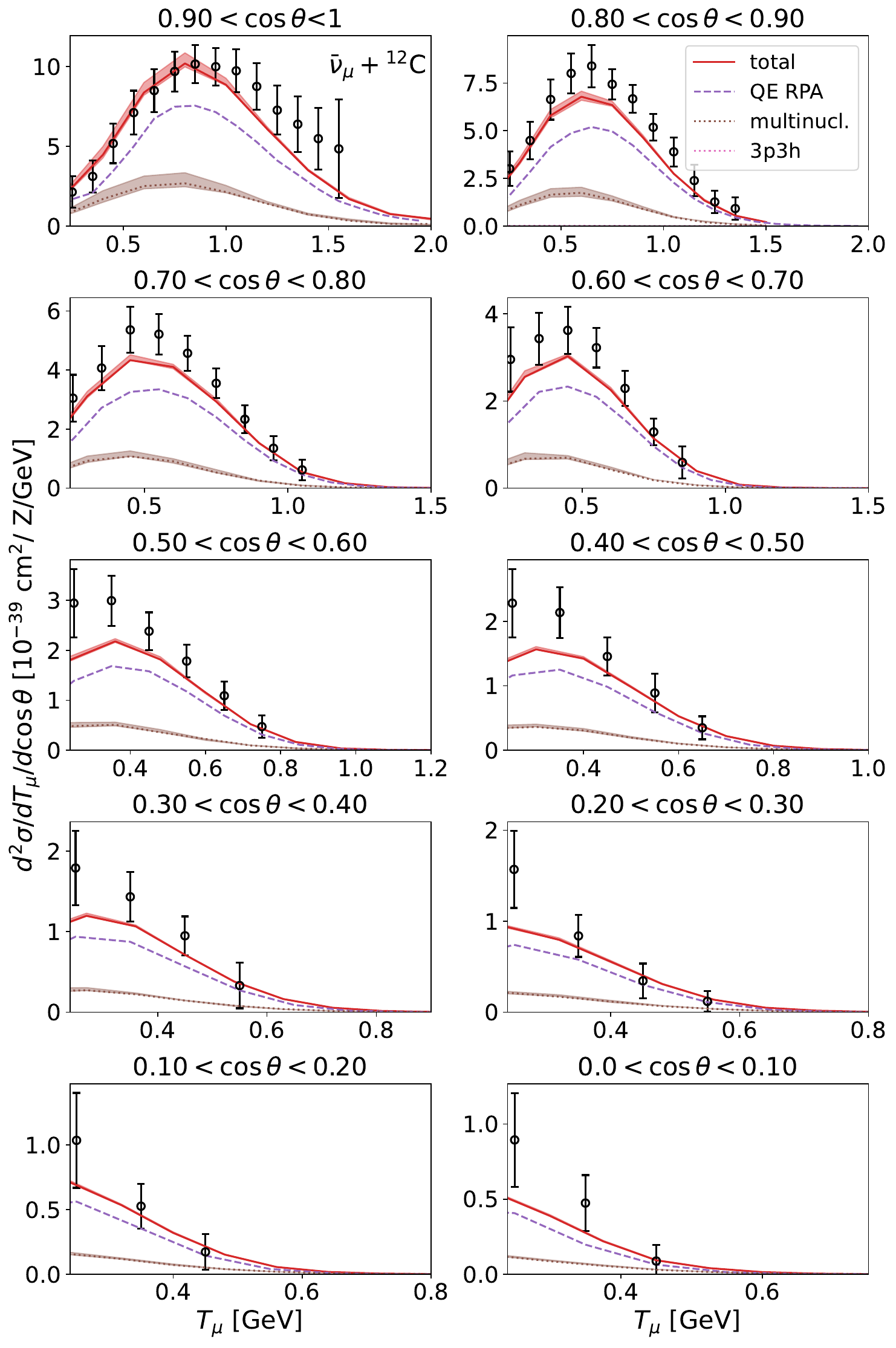}}
\caption{MiniBooNE neutrino~\cite{AguilarArevalo:2010zc} (left) and  antineutrino~\cite{MiniBooNE:2013qnd} (right) CCQE-like muon angle and kinetic energy distribution $d^2\sigma/d\cos\theta_\mu dT_\mu$ on carbon compared to predictions obtained in this work. Experimental errors do not include overall flux uncertainties.  We average the theoretical cross sections (integrate and divide by the length of the interval) for each of the $\cos\theta_\mu$-bins. The uncertainty bands in the theoretical results show  the variation produced when  ${\rm Re}\Sigma^{\rm eff}_\Delta(p_\Delta, q;\rho)$ change  in the range $[-50\rho/\rho_0, +30 \rho/\rho_0]$ MeV. The largest (smallest) cross sections are obtained for ${\rm Re}\Sigma^{\rm eff}_\Delta(p_\Delta, q;\rho)=-50\rho/\rho_0$ MeV ($+30\rho/\rho_0$ MeV), while the central values have been calculated using ${\rm Re}\Sigma^{\rm eff}_\Delta$=0.}
\label{fig:MB2D}
\end{figure}
\begin{figure}[tbh]
\centering
\makebox[0pt]{\includegraphics[height=.35\textwidth]{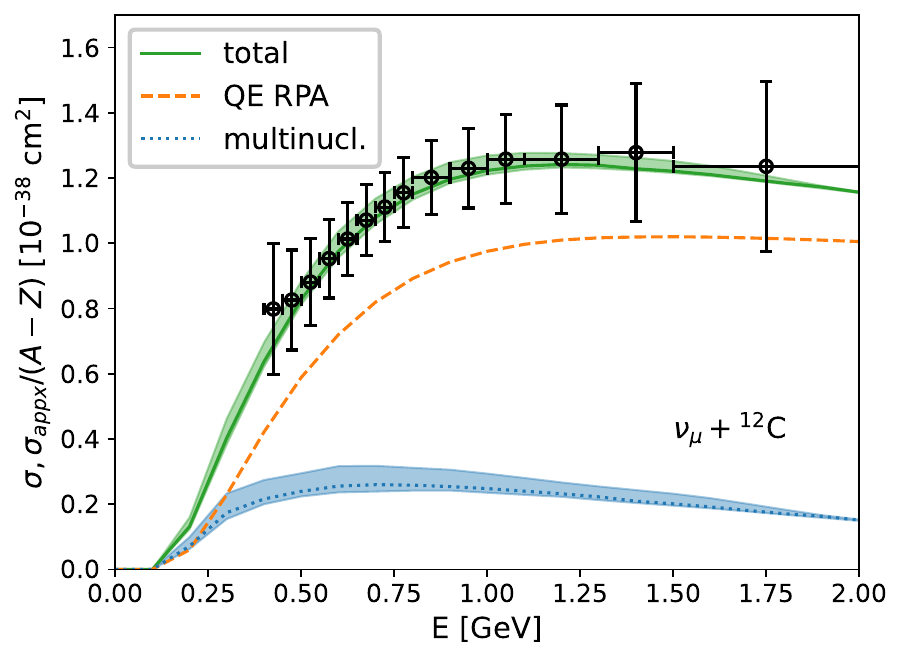}\hspace{0.75cm}\includegraphics[height=.35\textwidth]{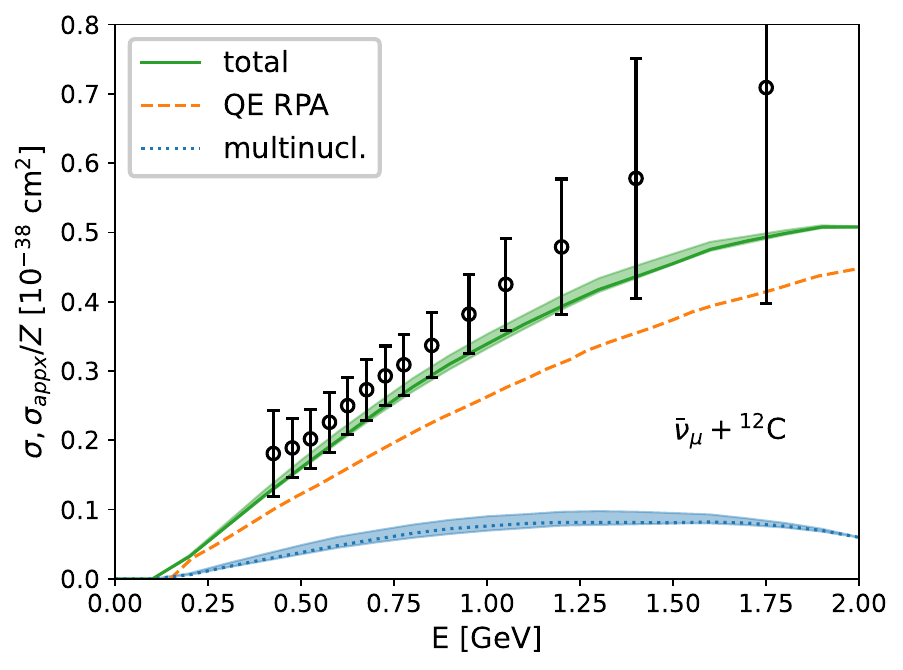}}
\caption{Flux-unfolded MiniBooNE muon neutrino~\cite{AguilarArevalo:2010zc} (left) and antineutrino~\cite{MiniBooNE:2013qnd} (right) CCQE-like cross section on carbon, divided by the number of active nucleons, as a function of (anti-)neutrino energy  compared to predictions obtained in this work. The theoretical multi-nucleon curves have been computed from the 2p2h cross section using $\sigma^{\rm 2p2h}_{\rm appx}$ defined in Eq.~(19) of Ref.~\cite{Nieves:2012yz}. The uncertainty bands in the theoretical results have been obtained as explained in Fig.~\ref{fig:MB2D}.  }
\label{fig:MBreco}
\end{figure}

\begin{figure}[h!]
\centering
\makebox[0pt]{\includegraphics[width=.375\textwidth]{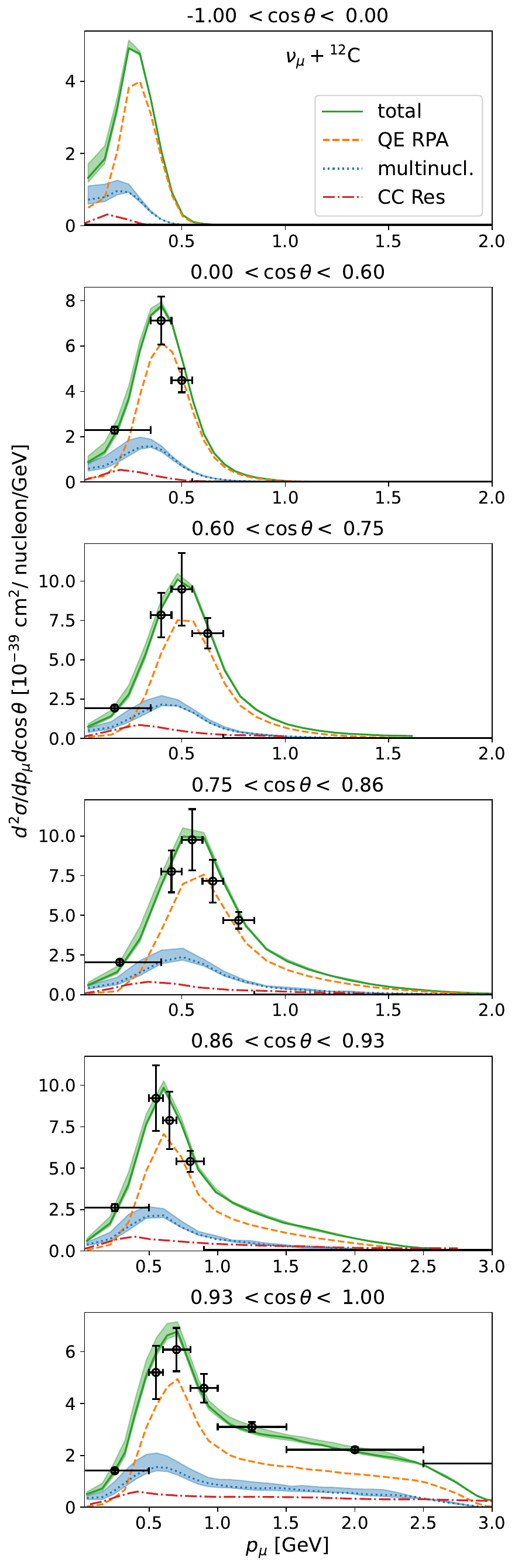}\hspace{0.5cm}\includegraphics[width=.38\textwidth]{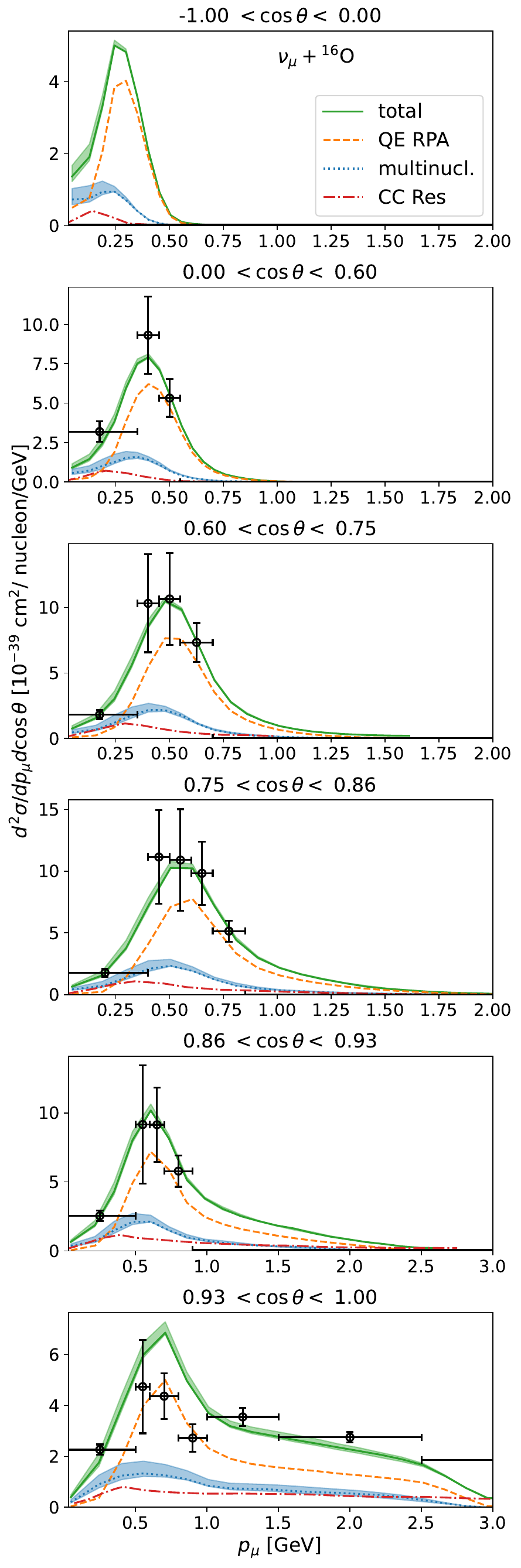}}
\caption{T2K neutrino  muon angle and momentum CC0$\pi$ double differential distribution $d^2\sigma/d\cos\theta_\mu dp_\mu$ on carbon (left) and oxygen (right) \cite{T2K:2020jav}, divided by the number of nucleons.  We also show the predictions obtained in this work,  supplemented by the contribution ``CC Res" taken from the NEUT results of Ref.~\cite{Bourguille:2020bvw} for carbon, and re-scaled by the number of nucleons for oxygen. We average the theoretical cross sections (integrate and divide by the length of the interval) for each of the $\cos\theta_\mu$-bins. The uncertainty bands in the theoretical results have been obtained as explained in Fig.~\ref{fig:MB2D}. }
\label{fig:T2K2D}
\end{figure}

Firstly, we compare the predictions of our updated model with the MiniBooNE CCQE-like neutrino and  antineutrino $d^2\sigma/d\cos\theta_\mu dT_\mu$ differential cross-section of Refs.~\cite{AguilarArevalo:2010zc,MiniBooNE:2013qnd}. We find an excellent reproduction of neutrino data across a broad range of muon kinetic energy and angular bins. In the earlier comparison of Ref.~\cite{Nieves:2011yp}, an agreement was found only when the neutrino flux was quenched by $10\%$, which could be accommodated thanks to the global normalization uncertainty of 10.7\% admitted by the MiniBooNE Collaboration~\cite{AguilarArevalo:2010zc}. The new 2p2h cross section  is larger than the one computed in \cite{Nieves:2011yp}, and allows for an excellent description of the experiment without having to reduce the data-points by an overall factor of 0.9.
Still, we support the findings of Ref.~\cite{Nieves:2011yp}, that the simultaneous consideration of RPA and multinucleon mechanisms makes the present microscopic model more appropriate than a pure impulse approximation. We also observe the significant size of the uncertainty on the theoretical 2p2h cross section induced by the variation of ${\rm Re}\Sigma^{\rm eff}_\Delta(p_\Delta, q;\rho)$ in the range $[-50\rho/\rho_0, +30 \rho/\rho_0]$ MeV (see discussion of item a) in Subsect.~\ref{sec:further}). The theoretical description of the MiniBooNE antineutrino CCQE-like double differential cross section (right panels of Fig.~\ref{fig:MB2D}) on carbon, though not as good as the neutrino one, is quite reasonable and certainly better than the predictions shown in \cite{Nieves:2013fr}. Here, we also confirm that the combined RPA and multinucleon nuclear effects lead to a robust microscopical description of the data. For all angular bins, multinucleon 
mechanisms improve agreement with data. Their contribution tends to concentrate at low muon energies so that more energy is transferred to the nucleus than in genuine QE events. Comparing antineutrino and neutrino double differential distributions, we observe a different pattern  because of the more forward peaked character of the antineutrino induced reaction and the lower average energies of the MiniBooNE antineutrino flux. For reference, the  3p3h contribution ($WNNN\to NNN$) is separately shown in the $0.80< \cos\theta_\mu<0.90$ window for both neutrino and antineutrino reactions. It is small in both cases because of the low energies of the MiniBooNE fluxes, and  in the antineutrino case, it is barely visible. 

Next, in Fig.~\ref{fig:MBreco} we show our predictions for  the flux-unfolded MiniBooNE muon neutrino~\cite{AguilarArevalo:2010zc} (left) and antineutrino~\cite{MiniBooNE:2013qnd} (right) CCQE-like cross section on carbon. The 2p2h contribution $\sigma^{\rm 2p2h}_{\rm appx}$ is calculated following the QE unfolding procedure explained in~\cite{Nieves:2012yz} (see Eq.~(19) of that work). The reproduction of the data for the neutrino reactions is spectacular, as good as that exhibited in Ref.~\cite{Nieves:2012yz} using the 2p2h contribution of Ref.~\cite{Nieves:2011pp}, but without relying on a global reduction of the data-points by a factor 0.9. Since in both works the genuine QE cross section is the same, the difference is only due to larger multinucleon cross sections here. For instance at $E=1$ GeV we obtain now, including the uncertainty due to the $\Delta$h RPA, values for $\sigma^{\rm 2p2h}_{\rm appx}/6$ (units of $10^{-38}$ cm$^2$)  in the range of  0.23-0.29,  while it was around 0.15 in  Ref.~\cite{Nieves:2012yz}. This quantitative difference does not alter the conclusions of Ref.~\cite{Nieves:2012yz} about the inadequacy of the genuine-QE algorithm to reconstruct the neutrino energy.  The MiniBooNE unfolded CCQE-like cross section exhibits an excess (deficit) at low (high) neutrino energies, which is an artifact
of an unfolding procedure that ignores multinucleon mechanisms. This effect should be now larger than that inferred from the solid green-curve and the pseudo data-points in Fig.~5 of Ref.~\cite{Nieves:2012yz}, since the 2p2h cross section is now bigger. This systematic bias is relevant to neutrino oscillation experiments that uses the CCQE-like samples to compute the neutrino energy.

The agreement with antineutrino unfolded cross-section on the right panel of Fig.~\ref{fig:MBreco} is good, actually it is slightly better than in  Ref.~\cite{Nieves:2013fr} where 2p2h results from Ref.~\cite{Nieves:2011pp} were used. For antineutrinos, we also find that the multinucleon cross section is now larger in this revised and corrected calculation, but the increase is significantly more moderate than in the case of the neutrino-induced reaction. Thus at $E=$1 GeV, $\sigma^{\rm 2p2h}_{\rm appx}/6$ takes values in the $(0.07-0.09)\times 10^{-38}$ cm$^2$  range, while one can read off $0.06 \times 10^{-38}$ cm$^2$ from the blue-circles of Fig.~1 of Ref.~\cite{Nieves:2013fr}. The error reported by MiniBooNE for the total cross sections is already of the order of  $0.04 \times 10^{-38}$ cm$^2$, comparable with the size of the unfolded multinucleon contribution, which makes multinucleon effects less relevant than in the neutrino driven process. 

We finish this part of the discussion of results with  Fig.~\ref{fig:T2K2D}, where we compare our predictions with the T2K neutrino  muon angle and momentum CC0$\pi$ double differential distributions  on carbon (left) and oxygen (right) \cite{T2K:2020jav}. Since these data-samples include all events with no pions in the final state, we have supplemented our QE+2p2h cross sections  by the small contribution  labelled as ``CC Res" in the plots and taken from the NEUT results of Ref.~\cite{Bourguille:2020bvw} for carbon. The ``CC Res'' contribution for oxygen has been obtained from carbon by re-scaling them by a factor $8/6$. It includes primarily events originated from production of a real pion in the initial step followed by its absorption before leaving the nuclear medium, which were removed by the MiniBooNE Collaboration from its two-dimensional distributions, as well as a very small deep inelastic scattering contribution from the higher-energy tail of the T2K neutrino spectrum.\footnote{We cannot extend our theoretical calculation of the QE and 2p2h cross sections to all T2K-flux neutrino energies and for these contributions we have  neglected the flux above 3 GeV.} For both nuclei, we reproduce quite well the distributions supporting the validity of the theoretical model discussed in this work. Results for carbon, using the 2p2h cross sections of Ref.~\cite{Nieves:2011pp}, are shown in the left panels of Fig.~5 of Ref.~\cite{Bourguille:2020bvw}, where it can be seen that a worse description of the data is found.

\subsection{\it Comparison with other theoretical calculations}

\begin{figure}[h]
\centering
\makebox[0pt]{\includegraphics[height=.35\textwidth]{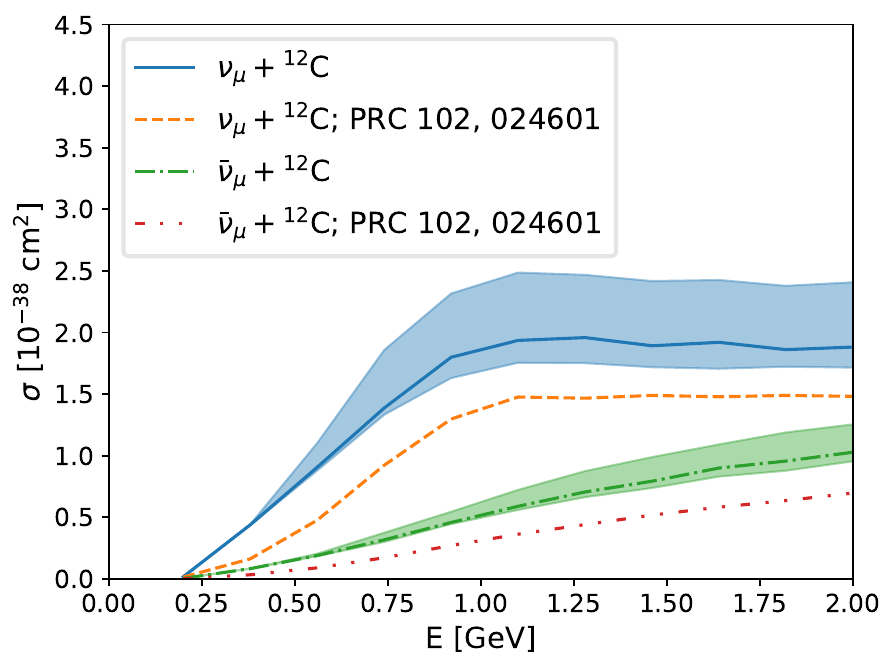}
}
\caption{Neutrino and antineutrino 2p2h (including three body absorption contributions) cross section  on $^{12}$C as a function of incoming neutrino energy $E$. We show both the results obtained in this work and those in Ref.~\cite{Sobczyk:2020dkn} (PRC 102, 024601). In all cases, a cut $|\vec{q}\,|<1.2$ GeV is implemented.  }
\label{fig:comps}
\end{figure}

In  Fig.~\ref{fig:comps}, we compare the actual neutrino and antineutrino 2p2h cross sections
$\sigma^{\rm 2p2h}$ evaluated in this work and those derived in Ref.~\cite{Sobczyk:2020dkn}. We clearly see the sizable effect produced by the inconsistent treatment of the energy conservation between the hole-nucleon and the virtual nucleon or $\Delta$ that enter in the $WN \to (N,\Delta)\pi$ vertices, which are part of the full $WN\to N'\pi$ amplitude. At $E=1$ GeV and in units of $10^{-38}$ cm$^2$, the new calculation provides for the neutrino mode a band of values comprised between 1.69 [for ${\rm Re}\Sigma^{\rm eff}_\Delta(p_\Delta, q;\rho)=+30\rho/\rho_0$ MeV] and 2.39 [for ${\rm Re}\Sigma^{\rm eff}_\Delta(p_\Delta, q;\rho)=-50\rho/\rho_0$ MeV], while the calculation carried out in Ref.~\cite{Sobczyk:2020dkn} gives around 1.38. These numbers for the antineutrino cross sections are [0.50-0.62] for the uncertainty band of the present evaluation and 0.31 for the calculation of  Ref.~\cite{Sobczyk:2020dkn}.

\begin{figure}[h]
\centering
\makebox[0pt]{\includegraphics[height=.35\textwidth]{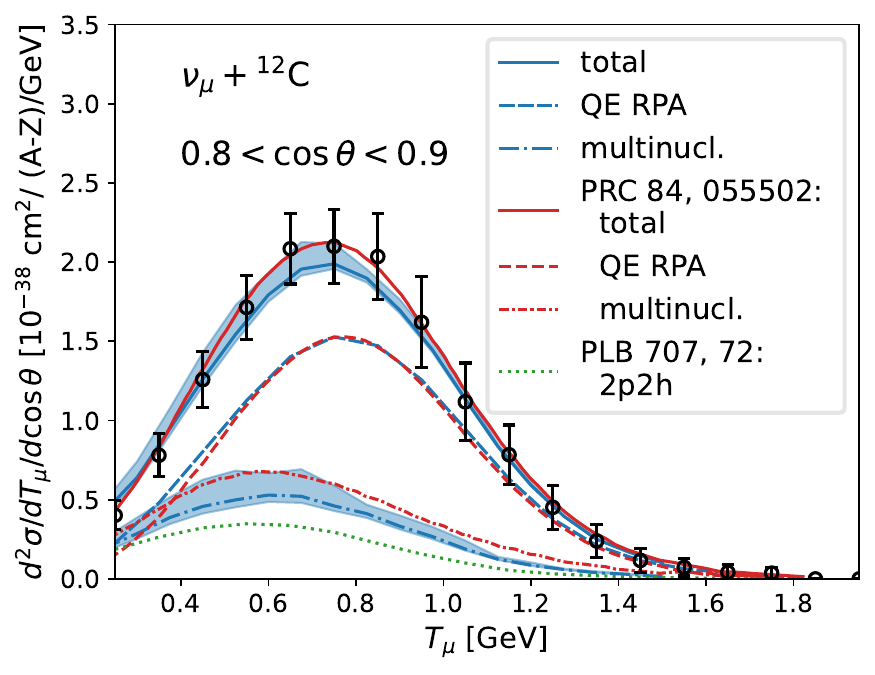}\hspace{0.5cm}
\includegraphics[height=.35\textwidth]{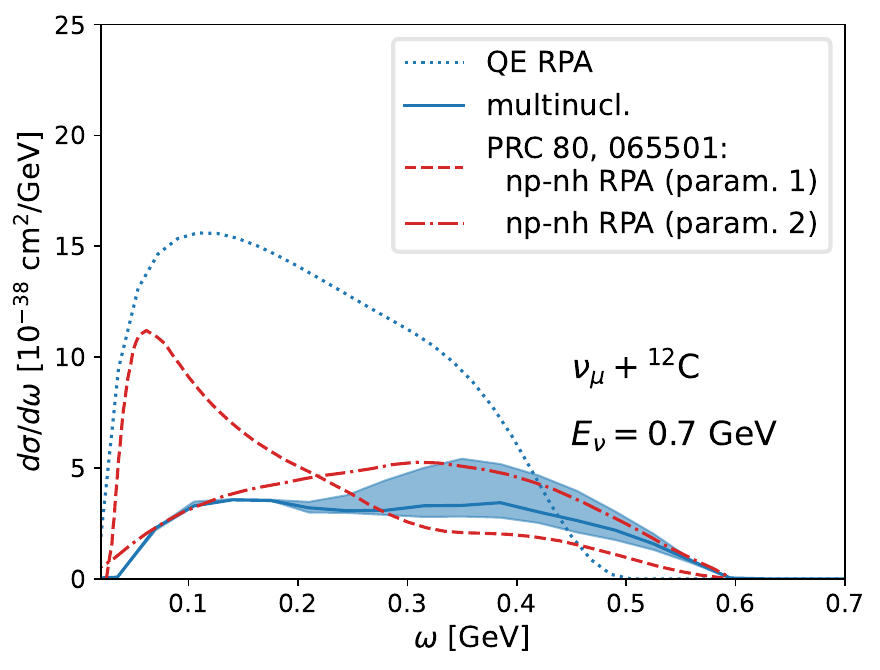}}
\caption{Left: MiniBooNE neutrino CCQE-like muon angle and kinetic energy distribution $d^2\sigma/d\cos\theta_\mu dT_\mu$ on carbon for $0.80 < \cos\theta_\mu < 0.90$ \cite{AguilarArevalo:2010zc}. We show separately the total, QE-RPA and multi-nucleon theoretical distributions obtained in this work, previously depicted in Fig.~\ref{fig:MB2D}, and by the Arpajon-Lyon group taken from Fig.~6 of Ref.~\cite{Martini:2011wp} (reddish curves, PRC 84, 055502). We also show the multi-nucleon contribution (green dotted curve, PLB 707, 72) from Fig.~3 of Ref.~\cite{Nieves:2011yp}. The uncertainty bands in the theoretical predictions of this work have been obtained in all cases as explained in Fig.~\ref{fig:MB2D}. 
Right: Various contributions to the CCQE-like $\nu_\mu+^{12}$C  differential cross section at $E=0.7$ GeV from this work, as a function of the energy transferred to the nucleus. Quasi-elastic calculations \cite{Nieves:2004wx} include RPA corrections and the uncertainty band induced by the variation of ${\rm Re}\Sigma^{\rm eff}_\Delta$ (see text of Fig.~\ref{fig:MB2D}) is shown for the multi-nucleon distribution. For this latter cross section, we also show the Lyon group results (red curves) of Fig.~22  of Ref.~\cite{Martini:2009uj} for the two different extrapolations  examined in that work. The one peaked at low values of $\omega$ is based on the approach of J. Marteau of Ref.~\cite{Marteau:1999kt}, while the another one exploits a microscopic evaluation of the 2p2h contribution to the transverse
magnetic response of $(e,e')$ scattering \cite{Alberico:1983zg}.  }
\label{fig:comps2}
\end{figure}

Lastly, we would like to compare our predictions with the Lyon group results~\cite{Martini:2009uj, Martini:2011wp}. Both models are based on the LDA, include RPA effects and have clear theoretical similarities in the treatment of the $\Delta$ in the medium.
The RPA 1p1h-1p1h, 1p1h-$\Delta$h and $\Delta$h-$\Delta$h forces are constructed qualitatively in a similar way, namely from $\pi+\rho+{\rm SRC}$ interactions, with similar parameters as in \cite{Nieves:2004wx}, except the $\rho$-driven potential which is a  factor 2/3 weaker in the Lyon group scheme.
The original calculations of Ref.~\cite{Martini:2009uj} are non-relativistic. In the later work of the Lyon group~\cite{Martini:2011wp}, a relativistic modification of the nuclear response is implemented as proposed in Refs.~\cite{Barbaro:1995ez,DePace:1998yx}. This amounts to work in a non-relativistic frame with the substitution $\omega \to \omega(1+\omega/2M)$ in the nuclear responses, and by multiplying the responses by $(1+\omega/M)$. As far as we understand, similar recipes are also used for the multinucleon cross sections also given in \cite{Martini:2009uj}.

The Lyon 2p2h predictions of Ref.~\cite{Martini:2009uj} in a first step relies on Marteau's model of Ref.~\cite{Marteau:1999kt}. There, and attending to Fig.~1 of that work,  the 2p2h cross section  is computed from the  first many-body diagram of Fig.~\ref{fig2:w_se}, considering only the NP, CNP, $\Delta$P and C$\Delta$P terms of the $WN\to N'\pi$ amplitude, and replacing the OPE by a more realistic interaction. The direct $\Delta$P-$\Delta$P contribution to the 2p2h cross section is considered as driven by a $\Delta$ self-energy diagram, which is taken into account through the parametrization of the in-medium $\Delta-$width of the Valencia group~\cite{Oset:1987re}. The rest of contributions that are not reducible to a modification of the $\Delta$ width are accounted  in two different ways, as we will describe below.

In the left panel of Fig.~\ref{fig:comps2}  we compare our new results with those previously published in Refs.~\cite{Martini:2011wp} and \cite{Nieves:2011yp}  for the MiniBooNE neutrino CCQE-like muon angle and kinetic energy distribution $d^2\sigma/d\cos\theta_\mu dT_\mu$ on carbon in the angular window $0.80 < \cos\theta_\mu < 0.90$.
QE cross sections, including RPA effects, appear to  agree very well as expected, except for the lowest muon kinetic energies, where some deviations,  although small, are visible.
For the multinucleon cross section, the differences are more visible. Our new predictions (dashed-dotted line)  are substantially larger than those of Ref.~\cite{Nieves:2011yp} (dotted-line), based on the findings of Ref.~\cite{Nieves:2011pp}. The uncertainties induced by the lack of a proper RPA re-summation of $\Delta$h excitations (band  depicted in the plot) is not sufficient to cover for the discrepancies between both sets of predictions. Thus for instance, for the muon kinetic energy $T_\mu$  of 0.6 GeV,  the result from Ref.~\cite{Nieves:2011yp} is $0.34 \times 10^{-38}$cm$^2$/GeV/$(A-Z)$, while the band stands for the interval $0.49-0.67$ 
in the same units.  Indeed, the present CCQE-like results describe well the MiniBooNE data without requiring a
global 90\% re-scaling down of the flux. As we have already pointed out, there exist several improvements implemented
here, including the correction for the inconsistent treatment of the energy conservation between the hole-nucleon and
the virtual nucleon or $\Delta$ that enter in the $WN\to N  \pi$ vertices, responsible for this enhancement of the cross section.

On the other hand, our results are now closer to those obtained by the Lyon group in~\cite{Martini:2011wp} (red dashed-dotted line in the plot) and they can be accommodated within the uncertainty band of our scheme. This is reassuring, taking into account the theoretical similarities of both calculations. The results presented in the left panel of Fig.~\ref{fig:comps2} are flux-folded, which prevents from a more in-depth comparison. Therefore, in the right panel we show the QE and 2p2h contributions to the differential cross-section $d\sigma/d\omega$ for a given neutrino energy. As mentioned, 
the Lyon group performed two different parametrizations for the the multi-nucleon distribution, in which the pieces that are not reducible to a modification of the $\Delta$ width are accounted for in two different ways. The authors of \cite{Marteau:1999kt} employ an extrapolation of the  two-body pion absorption at threshold carried out in Ref.~\cite{Faessler:1980ka}. Within this model, an accumulation of the 2p2h strength at low energy is produced (dashed line, ``param 1'', in the right panel of Fig.~\ref{fig:comps2}). 
Alternatively, the authors of Ref.~\cite{Martini:2009uj} presented results  in which they exploit a microscopic evaluation   of the 2p2h contribution to the transverse
magnetic response of $(e,e')$ scattering \cite{Alberico:1983zg}. It improves on the previous parametrization which had no
momentum dependence, and it agrees better with our microscopic results (dashed-dotted line, ``param 2''). The Lyon group results shown in the left plot of Fig.~\ref{fig:comps2} were obtained using this more realistic model for the 2p2h contribution. 
Still, it becomes clear from the above discussion that neither of the two 2p2h models discussed in Ref.~\cite{Martini:2009uj} can be directly used to predict exclusive two-nucleon final states~\cite{Martini:2009uj,Martini:2010ex,Martini:2011wp,Martini:2013sha}.

%
%
%
% CONCLUSIONS
%
%
%

\section{Conclusions}
\label{sec:concl}
The ambitious experimental program which is underway to precisely determine neutrino properties, requires a solid microscopic understanding of neutrino-nucleus interactions. The (anti-)neutrino CC  multi-nucleon cross section calculation presented in this work provides an adequate theoretical tool for neutrino oscillation experiments at low and moderate  energies. It is based on a state of the art description of the neutrino pion production off the nucleon and a vast experience in describing nuclear responses sensitive to the spin-isospin channel of the $NN$, $N\Delta$ and $\Delta\Delta$ in medium interactions.   

We presented a revisited calculation of the CC multinucleon knockout process improved on several aspects of the computation of the  CCQE-like cross section carried out in Ref.~\cite{Sobczyk:2020dkn}. Importantly, we introduced a consistent treatment of the nucleon self-energy in the $WN\to N'\pi$ amplitude leading around $20-40\%$ enhancement in respect to the previous results.
We have taken the opportunity of this revision to discuss in  detail several important issues of the calculation of the 2p2h cross section, delving into the microscopic dynamics of the multi-nucleon mechanisms. Within this discussion, we identified the RPA series of $\Delta$h excitations driven by the $\pi+\rho$+SRC spin-isospin interaction, as the largest source of theoretical uncertainty on the 2p2h cross sections presented in this work.

Our current predictions agree very well at the inclusive level with the (anti-)neutrino CCQE-like and neutrino CC0$\pi$ lepton double-differential cross sections measured by  MiniBooNE~\cite{AguilarArevalo:2010zc,MiniBooNE:2013qnd} on carbon and by T2K~\cite{T2K:2020jav} on carbon and oxygen, respectively. In particular,  the neutrino MiniBooNE data is now well described without requiring a global 90\% re-scaling down of the flux, as it was needed in Ref.~\cite{Nieves:2011yp}.

Moreover, as pointed out in \cite{Sobczyk:2020dkn}, the microscopic calculation performed in our scheme provides the two-nucleon final states in the primary vertex of the interaction. 
Work is currently underway to include our predictions in Monte Carlo event generators. It will allow to account for the exclusive distributions of the outgoing nucleons, modified on their way leaving out the nucleus. Consequently, we will be able to make a comparison of our predictions, including both the QE and multinucleon knockout mechanisms, with semi-exclusive cross-sections measured with an increasing statistics by present (e.g. T2K, MicroBooNE) and future (e.g. SBND) experiments.

\section*{Acknowledgements}
We warmly thank Jos\'e Enrique Amaro, Eulogio Oset, Federico S\'anchez and Jan Sobczyk for useful discussions and clarifications.  
J.E.S. acknowledges the support of the European Union’s Horizon 2020 research and innovation programme under the Marie Skłodowska-Curie grant agreement No. 101026014.
This research has been supported  by the Spanish Ministerio de Ciencia e Innovaci\'on (MICINN)
and the European Regional Development Fund (ERDF) under contract PID2020-112777GB-I00, the EU STRONG-2020 project under the program H2020-INFRAIA-2018-1,  grant agreement no. 824093. It has also been
funded by  Generalitat Valenciana under contract PROMETEO/2020/023 and the “Planes Complementarios de I+D+i” program
(Grant No. ASFAE/2022/022) by MICINN with funding from
the European Union NextGenerationEU and Generalitat
Valenciana, the Deutsche
Forschungsgemeinschaft (DFG)
through the Cluster of Excellence ``Precision Physics, Fundamental
Interactions, and Structure of Matter" (PRISMA$^+$ EXC 2118/1) funded by the
DFG within the German Excellence Strategy (Project ID 39083149).

\bibliography{neutrinos}

\end{document}